# Melting of San Carlos olivine in the presence of carbon at 6-12 GPa

**Jozsef Garai* and Tibor Gasparik**
Department of Geosciences
State University of New York at Stony Brook
Stony Brook, New York, 11794, USA

Corresponding Author: Jozsef Garai
E-mail: Jozsef.Garai@fiu.edu
Tel: 305-348-3445
Fax: 305-348-3070
*Current Address:
Department of Earth Sciences
Florida International University
University Park, PC 344
Miami, Florida 33199, USA

**Abstract** Experiments at 6-12 GPa showed that San Carlos olivine surrounded by carbon melts incongruently at 1400 $^{o}C$ and higher temperatures, producing olivine with lower Fe contents, pyroxene, carbide, and carbonate melt. The relatively low melting temperature of 1400 $^{o}C$, independent of pressure, is consistent with carbonate melting. The new evidence for the reduced stability of olivine at high temperatures in the presence of carbonate melts is consistent with geophysical observations, and results in a substantially improved agreement between the experimentally determined $(Mg, Fe)_2SiO_4$ phase relations and the observed seismic velocity structure of the upper mantle.

## Introduction

Many believe that olivine and other $(Mg, Fe)_2SiO_4$ polymorphs stable at higher pressures, wadsleyite and ringwoodite, are the dominant mineral constituents of the upper mantle. The experimentally determined solidus of the anhydrous peridotite KLB-1 (Herzberg 2000; Zhang 1994) relevant for such a mantle, is hundreds of degrees higher than the expected temperatures of an average mantle. In contrast, the presence of partial melt in the low velocity zone (Gutenberg 1948) and possibly in the depth range of 500-1000 km (Cadek 1998; Kido 1997; Montagner 1998) is suggested by the detected high attenuation of seismic waves. Partial melting of the KLB-1 peridotite was observed in the same experimental study (Herzberg 2000) even below the solidus temperatures, when graphite containers were used instead of the rhenium capsules, thus suggesting that the presence of graphite could lower the melting temperatures. Carbon is known to exist in the mantle and thus could affect the stability and melting temperature of olivine.

## Experimental

### Diamond Aggregate

In order to obtain new constraints on the properties of an olivine-rich mantle at high pressures and temperatures, we used multianvil presses to carry out experiments at 6-12 GPa with single crystals of San Carlos olivine surrounded by a matrix of diamonds (25-80 $\mu m$) enclosed in rhenium capsules. The experimental conditions are listed in Tab. 1. The equipment, the 10-mm sample assembly used for the experiments, temperature and pressure calibrations and experimental procedures were described elsewhere (Gasparik 1989; Li 1996).

When the experiments with diamonds lasting from 20 minutes to 4 hours were completed, the experimental products were removed from the capsules, and the diamond matrix was burned off in a furnace at $800\,^{\circ}C$ in 2 days to ease the preparation of the products for optical observation, microprobe analysis and transmission electron microscopy (TEM).

Graphite

A diamond aggregate has been used before to collect melt in other melting studies (Hirose 1993). This technique is sometimes criticized because the pressure in the sample may not be uniform. To avoid this potential problem, we assumed that graphite could provide the same chemical environment as diamond, and that the applied stress on the crystal fragment should be close to hydrostatic. Using graphite had an additional advantage that the preparation of thin sections would not require burning. Experiments with fragments of San Carlos olivine surrounded by graphite were carried out at 6 and 12 GPa and 1400 $^{o}C$, and at 12 GPa and 1300 $^{o}C$.

**Results**

Diamond Aggregate

Minimal or no changes to the fragments were observed in the experimental products obtained at temperatures below 1400$^{o}C$. However, significant changes were noted at 1400 $^{o}C$ and higher temperatures: The original fragments were surrounded by fine red whiskers, which were apparently present in the space between the diamonds before the burning. Thin sections showed that the original single crystal olivine had completely recrystallized to a polycrystalline mosaic of grains with an average grain size of 50 $\mu m$. In the experiment carried out at 6 GPa and 1180 $^{o}C$, the red whiskers were not present, and the original single crystal did not recrystallize.

The original San Carlos olivine and the experimental products were analyzed by electron microprobe. The average compositions of phases are listed in Tab. 1. The $Fe_2SiO_4$ content of the fragments from the experiments carried out at 6 GPa/1400 $^{o}C$, and 12 GPa/1500 $^{o}C$, was about 5-mol %, which is about 4% less than the original composition of $9.09(\pm 0.31)\%$. There was no detectable change in the Fe content of the olivine fragment from an experiment at 10 GPa and 1400 $^{o}C$. The material transported into the diamond aggregate was analyzed in a sample obtained at 12 GPa and 1500 $^{o}C$. The average sum of the cations was 4.48 per 6 oxygens, thus showing a stoichiometry close to olivine. The average $Fe_2SiO_4$ content was 11.2 mol %, which is about 6 % higher than in the olivine fragment from the same experiment.

In order to make observations at higher resolution from the sample obtained at 6 GPa and 1400 $^{o}C$, the edge of the original fragment, which had been in contact with the diamond aggregate, was investigated with TEM equipped with an energy dispersive X-ray spectrometer. Quenched Fe rich melt was detected in the triple junctions and along the grain boundaries (Fig. 1.). Dihedral angles were measured on several TEM images (e.g. Fig. 1.), and were found to be in the range of 0-35$^{o}$. These low dihedral angles indicate that the melt in the triple junctions was interconnected. The olivine grains were generally free of dislocations, indicating a low differential stress in olivine. We used the Tracor EDX Detector of a Jeol 200 CX microscope to obtain the qualitative composition of the quenched melt, and a typical compositional spectrum is given in Fig. 2. The spectrum shows that the quenched melt was extremely rich in Fe with some minor Mg. The virtual absence of Si indicates that the potential overlap with the surrounding olivine was minimal, and that the analysis was representative of the melt composition. The EDX detector in use, unfortunately, does not allow the detection of light elements, such as C or O. Hence, from the compositional spectrum alone, we were not able to decide what kind of melt was present in the triple junctions and at the grain boundaries. We also tried to obtain structural information from some hexagonal looking crystals located within the quenched melt by using selected area electron diffraction analysis on a single grain. The grain diameter of the crystals at the grain boundaries was very small, typically around 100 nm. This diameter is smaller than the smallest Selected Area (SA) aperture (~0.4 $\mu m$ in the sample image plane) that equipped the microscope. However, we were able to obtain one single-crystal diffraction pattern (Fig. 3.). The diffraction pattern is partly polluted by diffraction spots from surrounding crystals. To interpret this diffraction pattern, we used the Electron Microscope Image Simulation (http:et al. ). Several compounds, with hexagonal or orthorhombic structure, including pure iron, $FeCO_3$, and most of the iron carbides, $Fe_2C$, $Fe_3C$, $Fe_7C_3$ were tested. We interpret the single crystal diffraction pattern as either $Fe_3C$, group symmetry Pnma62, orthorhombic with the [11,3,1] zone axis, or $Fe_7C_3$, group symmetry Pnma62, orthorhombic, with the [4,-19,6] zone axis. We also obtained a powder diffraction pattern which indicated that the hexagonal looking crystals observed at grain boundaries were most likely $Fe_3C$ carbide.

Graphite

In the experiments carried out at 1400 $^{o}C$, full recrystallization of the original single crystal fragment was again observed, producing crystals with an average grain size of $50\mu m$. The TEM investigation of the experimental product obtained at 6 GPa and 1400 $^{o}C$ did not find any melt in the recrystallized olivine. Microprobe analyses did not show a measurable change in the Fe content of olivine in the samples after the experiments in comparison with the original San Carlos olivine. However, the presence of Fe, Mg, and Si was again detectable within graphite by electron microprobe. These observations suggested that the chemical reactions and the transport of material from the original fragments of San Carlos olivine also occurred during the experiments with graphite. In contrast, no change to the original single crystal of San Carlos olivine was observed after an experiment at 12 GPa and 1300 $^{o}C$.

In order to obtain a measurable quantity of the transported material in graphite, the run duration was extended from 20-25 min to 12 hours in an experiment carried out at 6 GPa and 1400 $^{o}C$. In this experiment, small crystals dispersed in graphite were visually detected under a standard optical microscope (Fig. 4). The original Fe/(Fe+Mg) atomic percentage of olivine was reduced from $9.09 \pm 0.31$ % to $8.51 \pm 0.15$ %, and the crystals within graphite were identified as garnet, pyroxene, spinel, and possibly quenched melt. The clearly identified presence of garnet in graphite suggested that garnet might have been present also in the whiskers in the experiments with diamond, particularly in the experiment at 12 GPa and 1500 $^{o}C$. In this experiment, the electron microprobe analyses of four points showed a stoichiometry close to 4.0 per 6 oxygens, and the aluminum content of 0.3-0.7 per 6 oxygens. However, the quality of these analyses was very poor due to the small size of the whiskers.

The microprobe and TEM investigations of the original single crystal did not detect anything else beside olivine. In order to determine the source of the Al the original single crystal olivine was investigated by x-ray mapping. The detected local maxima in the distribution of Al suggest that the San Carlos olivine contained spinel micro inclusions. The size of this inclusion has to be below the resolution of TEM. The precipitation of Al-rich spinel and enstatite in San Carlos olivine at 4.6-9.0 GPa and 1310-1595 $^{o}C$ was also reported (Raterron 1998).

**Discussion**

The observed changes in the experimental products, such as the recrystallization, the reduced Fe content of the fragments, and the transport of material into diamond or graphite, suggested that the fragments of San Carlos olivine underwent partial melting. The melting temperature of 1400 $^{o}C$, observed in our experiments at 6 and 12 GPa, seems to be independent of pressure, which is consistent with carbonate melting. In an experimental study of melting of the primitive kimberlite Jd-69, the solidus temperature increased only slightly from 1380 to 1430 $^{o}C$ in the pressure range of 6-12 GPa (Wang 1999).

We interpret the observed changes in the following way. The presence of carbon imposed a low oxygen fugacity in the system and olivine melted incongruently, producing olivine with lower Fe content, pyroxene, carbide and carbonate melt. Pyroxene also reacted with spinel and produced garnet. In the run products from the experiments with diamond, two distinct occurrences of melt were observed, one in the triple junctions along the edge of the fragment, and the other in whiskers located within the diamond aggregate. Apparently, the melt, which produced the whiskers, was transported from the fragment in response to the pressure gradients resulting from the high strength of the diamond aggregate. In the experiments with graphite, pockets of quenched melt were located near the interface between the fragment and graphite. The chemical reaction, occurred at 6 GPa and 1400 $^{O}C$, is:

$$18Fe_2SiO_4 + 10C = 9Fe_2Si_2O_6 + 6FeCO_3(\text{melt}) + 4Fe_3C \quad (1)$$

These results are consistent with earlier studies which showed that in the pressure and temperature range of our experiments under $CO_2$–saturated conditions, $CO_2$ would react with forsterite to produce enstatite and magnesite (Katsura 1990; Koziol 1998; Newton 1975). If enough $CO_2$ is supplied, olivine should completely decompose.

In our experiments, only a very small amount of olivine was decomposed, indicating thus that the cold sealing of the capsules employed in our experiments was sufficient to provide a closed system. The redox conditions, applied during the experiments, were determined by the buffering effects of olivine, graphite, and diamond. It has been proposed that C-O fluids in equilibrium with elemental carbon buffer the oxidation state of the earth's upper mantle (French 1966; Sato 1978). The observed higher percentage of melt in the experiments with diamond could be the effect of more oxygen in the air trapped among the diamond grains; therefore, we consider the conditions of graphite experiments are relevant to the mantle.

The quantity of melt ($FeCO_3 + Fe_3C$) produced in our experiments with diamond at 6 GPa was estimated to be equivalent to 3.0 volume % of the original fragment, using the largest observed change in the Fe/(Fe+Mg) ratio of olivine and the proposed chemical reaction in the calculations. The approximate volume ratio between the carbonate melt and the carbide melt is 2 to 1. At higher pressures, 10 and 12 GPa only $FeCO_3$ should be present as melt, while $Fe_3C$ or $Fe_7C_3$ would occur in the melt as crystals. Using the same method, the calculated amount of melt in the experiments with graphite would be around 0.5 volume %.

In the presence of carbon as carbonate melt, olivine in the mantle should have a lower Fe/(Fe+Mg) atomic ratio than the San Carlos olivine, as indicated by our experiments. This prediction is consistent with observations. In the melting of the primitive aphanitic kimberlite (JD 69), 5.5-7.5 mol % of $Fe_2SiO_4$ was reported in olivine (Wang 1999). These values are also in agreement with the 7-8% in olivine found as inclusions in diamonds and in xenoliths from kimberlite (Meyer 1972). Based on the Fe content of olivine observed as inclusions in diamonds, about one percent of melt can be expected in the upper mantle.

The presence of carbonate melt modifies the subsolidus phase relations for $(Mg, Fe)_2SiO_4$. Using the Fe/(Fe+Mg) atomic ratios observed in kimberlites and diamond inclusions the solidus of olivine coexisting with carbon is given in Fig. 5. In regions where carbonate melt is present in the mantle, the carbonate solidus should be used for the mantle instead of the silicate solidus. A temperature-pressure phase diagram for a mantle with the composition $(Mg_{0.9}Fe_{0.1})_2SiO_4$ is shown in Fig. 6. This solidus suggests that carbonate melt should be present in the upper mantle down to the transition zone, and perhaps even deeper than 660 km. Carbonate melts could separate through geological times and become trapped in higher concentrations at the bottom of the lithosphere and perhaps at the bottom of the 660 km boundary, as suggested by the observed seismic wave velocity attenuation.

The existence of carbonate melt in the mantle is supported by the line of direct and indirect evidences (Green 1988; Harmer 1998; Hauri 1993; Rudnick 1993; Yaxley 1991). While the presence of the carbonate melt in the mantle has not been questioned, its origin is still debated. The incongruent melting of olivine occurring in the presence of carbon, as detected in our experiments, offers a plausible explanation for the origin of the carbonate melt in the mantle. Carbon is known to exist in the mantle as graphite or diamond, which shows that free carbon still available after the production of carbonate melt. Thus the amount of carbon is/was sufficient to produce the proportion of the carbonate melt observed in our experiments.

**Acknowledgements**

We thank P. Raterron for conducting the TEM investigations, P. Raterron and D. H. Lindsley for useful discussions and Florida International University, Florida Center for Analytical Microscopy for the x-ray mapping.

**TABLE 1.** Experimental conditions and average compositions of phases

| Run | t (min.) | P (GPa) | T (°C) | Location | | Points Analyzed | Cations/6 Oxygens Fe | Mg | Si | Al | Ca | Sum | Weight Totals | 100Fe/(Fe+Mg) |
|---|---|---|---|---|---|---|---|---|---|---|---|---|---|---|
| K-546 | 660 | 6 | 1,400 | fragment, middle | Ol | 12 | 0.255 | 2.746 | 1.497 | 0.001 | 0.003 | 4.502 | 98.0 | 8.50 |
| (Graphite) | | | | edge | Ol | 5 | 0.235 | 2.780 | 1.491 | 0.000 | 0.003 | 4.509 | 98.2 | 7.79 |
| | | | | in the graphite | Ga | 11 | 0.225 | 1.195 | 1.507 | 0.985 | 0.084 | 3.996 | 96.0 | 15.85 |
| | | | | | S | 4 | 0.023 | 2.605 | 1.110 | 0.777 | 0.001 | 4.516 | 95.3 | 0.90 |
| | | | | | L | 2 | 0.258 | 4.033 | 0.853 | 0.000 | 0.003 | 5.147 | 69.9 | 6.05 |
| | | | | | Ol | 3 | 0.217 | 2.723 | 1.522 | 0.008 | 0.004 | 4.474 | 96.9 | 7.38 |
| | | | | | Px | 2 | 0.119 | 1.784 | 1.970 | 0.099 | 0.010 | 3.982 | 96.2 | 6.26 |
| K-535 | 25 | 12 | 1,400 | fragment, middle | Ol | 11 | 0.280 | 2.739 | 1.484 | 0.003 | 0.004 | 4.510 | 98.7 | 9.27 |
| (Graphite) | | | | | Ga | 1 | 0.373 | 1.061 | 1.416 | 1.106 | 0.055 | 4.011 | 93.7 | 26.00 |
| | | | | edge | Ol | 9 | 0.270 | 2.732 | 1.477 | 0.024 | 0.003 | 4.506 | 97.7 | 9.01 |
| K-542 | 20 | 12 | 1,300 | fragment, middle | Ol | 10 | 0.272 | 2.744 | 1.489 | 0.001 | 0.004 | 4.510 | 97.8 | 9.01 |
| (Graphite) | | | | edge | Ol | 3 | 0.272 | 2.737 | 1.492 | 0.001 | 0.004 | 4.506 | 98.3 | 9.04 |
| S-3316 | 240 | 12 | 1,500 | fragment, middle | Ol | 4 | 0.158 | 2.886 | 1.476 | 0.001 | - | 4.521 | 99.6 | 5.20 |
| (Diamond) | | | | edge | Px or L | 1 | 0.170 | 2.405 | 1.708 | 0.004 | - | 4.287 | 99.3 | 6.59 |
| | | | | whiskers | L | 11 | 0.322 | 2.577 | 1.533 | 0.012 | - | 4.444 | 31.5 | 11.11 |
| | | | | | Ga | 3 | 0.247 | 1.426 | 1.656 | 0.649 | - | 3.978 | 28.5 | 15.24 |
| | | | | | Px | 1 | 0.203 | 1.463 | 2.153 | 0.004 | - | 3.823 | 14.5 | 12.20 |
| S-3413 | 20 | 10 | 1,400 | fragment, middle | Ol | 9 | 0.276 | 2.752 | 1.481 | 0.002 | 0.004 | 4.515 | 98.0 | 9.09 |
| (Diamond) | | | | edge | Ol | 6 | 0.240 | 2.720 | 1.497 | 0.026 | 0.003 | 4.486 | 98.9 | 8.09 |
| | | | | | L | 1 | 0.290 | 2.636 | 1.438 | 0.128 | 0.005 | 4.497 | 96.5 | 9.90 |
| K-474 | 30 | 12 | 1,400 | fragment, middle | Ol | 4 | 0.266 | 2.713 | 1.507 | 0.001 | 0.004 | 4.491 | 97.5 | 8.93 |
| (Diamond) | | | | edge | Ol | 2 | 0.258 | 2.708 | 1.514 | 0.001 | 0.004 | 4.485 | 97.4 | 8.69 |
| **Original San Carlos Olivine** | | | | | | **16** | **0.273** | **2.735** | **1.491** | **0.002** | **0.004** | **4.505** | **98.2** | **9.09** |

Ol = Olivine
Px = Pyroxene
Ga = Garnet
S = Spinel
L = Liquid

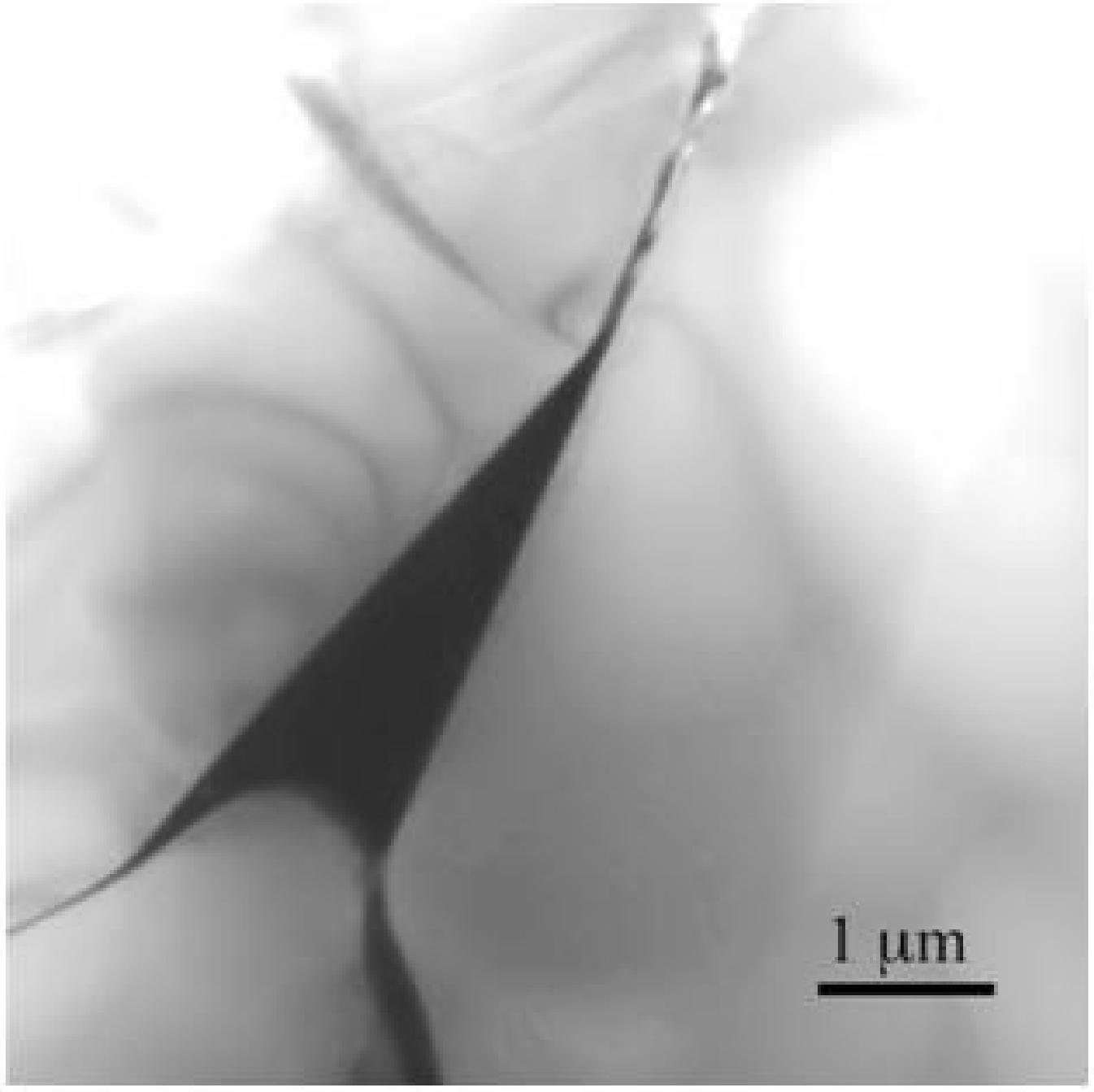

**Fig. 1** Brightfield TEM image of a triple junction between olivine grains in the sample obtained at 6 GPa and 1400 , showing quenched Fe-rich melt in the triple junction and along the grain boundaries. (Investigator P. Raterron)

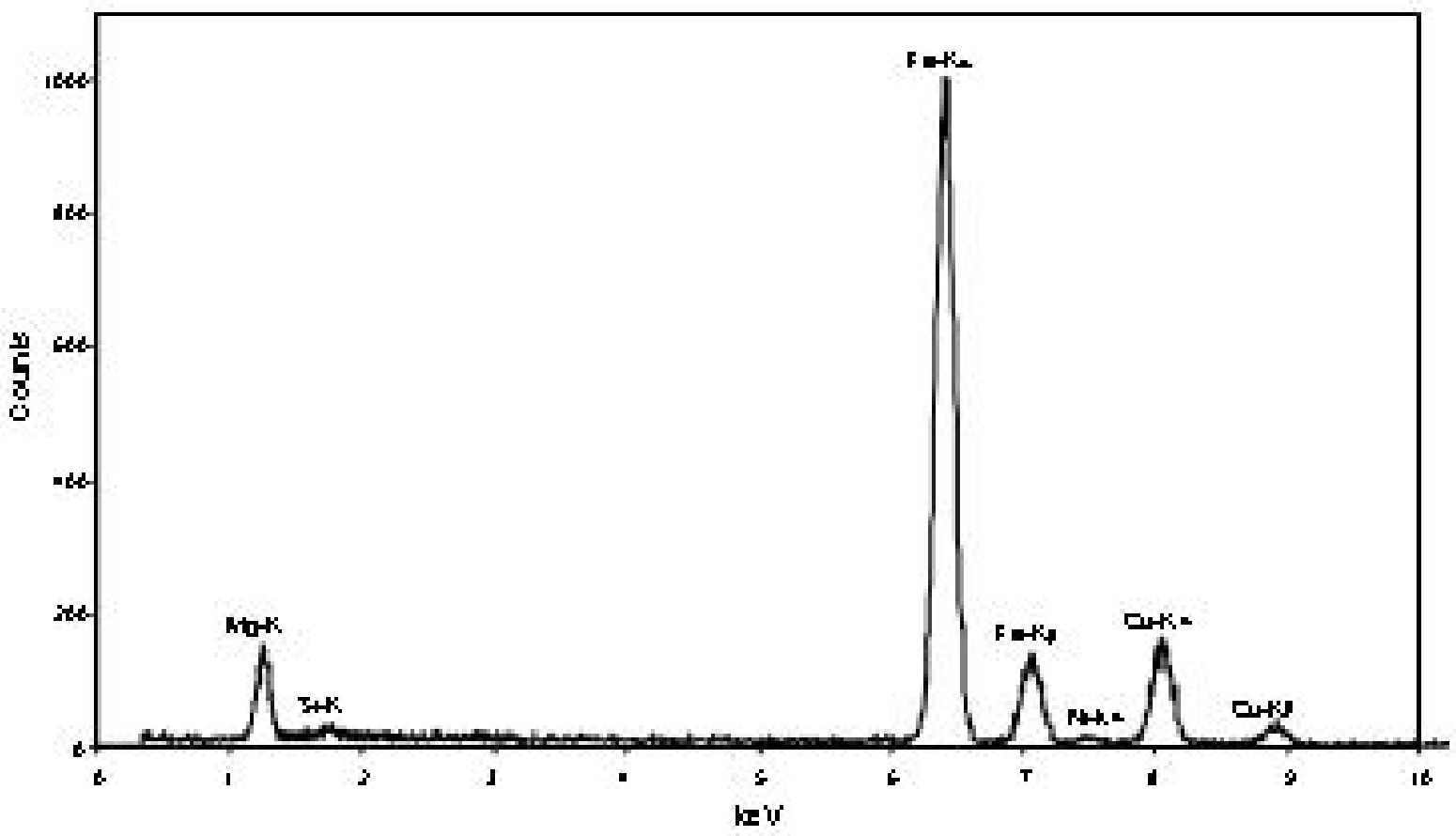

**Fig. 2** Energy dispersive X-ray spectrum of the Fe-rich melt from the triple junction shown in Fig. 2, obtained in the TEM mode by focusing the electron beam on a thin area of the specimen (less than 100-nm thick). Light elements, such as oxygen and carbon, cannot be detected through the beryllium window that equips the detector. The copper peaks result from the sample mount and are not characteristic of the analyzed material. (Investigator P. Raterron)

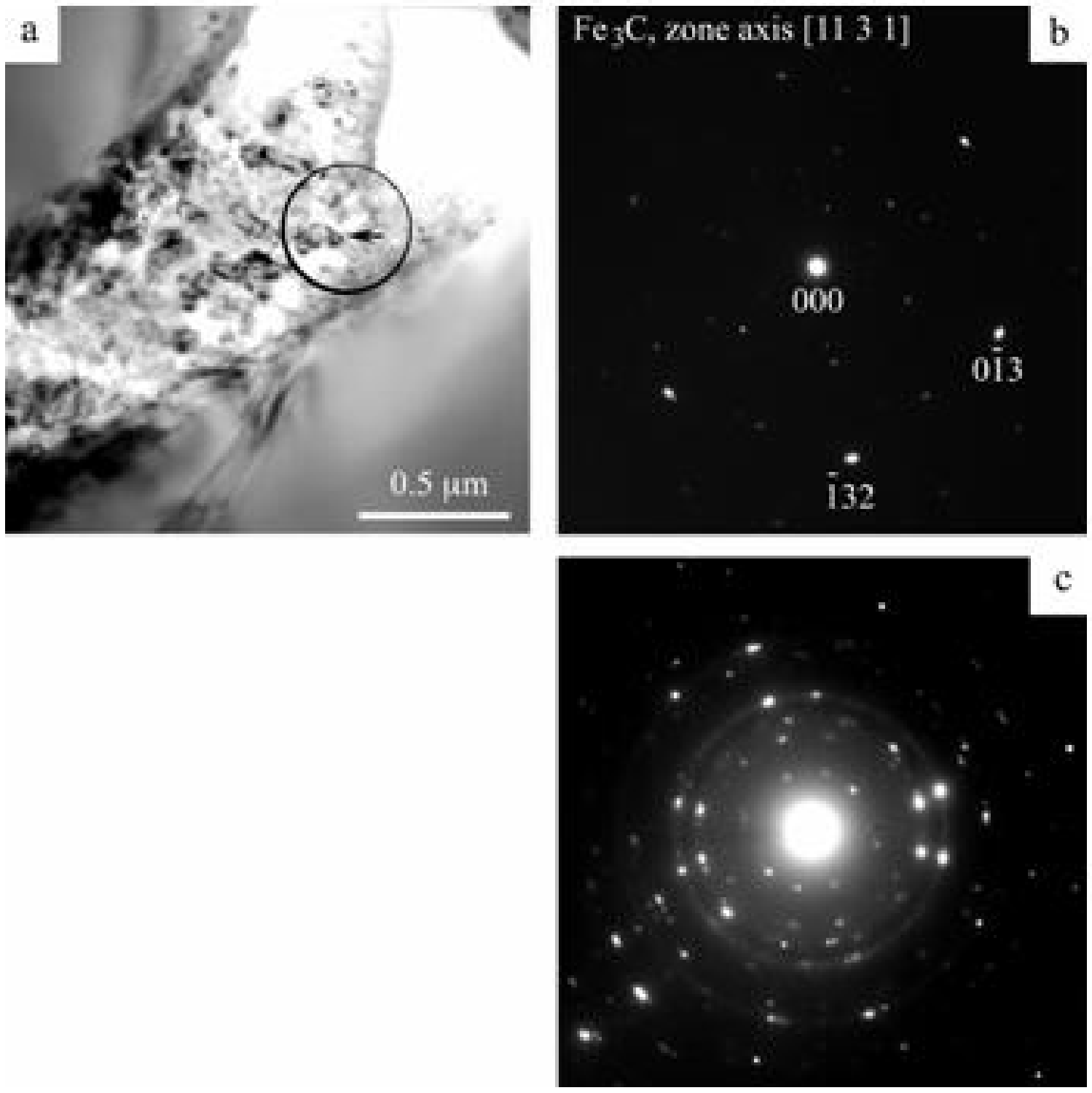

**Fig. 3** TEM investigation of a triple junction in the sample obtained at 6 GPa and 1400 . (Investigator P. Raterron)

**a.**/ Bright field TEM micrograph of the quenched melt. Individual crystals (less than 100 nm in size) are visible. The black circle shows the approximate size and position of the Selected Area (SA) aperture used to tentatively identify the crystal marked by an arrow.

**b.**/ TEM Selected Area diffraction pattern obtained from the crystal showed in the previous micrograph by the arrow. Despite the presence of a few weak extra spots, probably due to other diffracting grains, this pattern corresponds to the Fe3C [11, 3, 1] zone-axis pattern.

**c.**/ TEM Selected Area diffraction pattern obtained from a thicker area, containing many crystals, of the same triple junction. This pattern shows the superposition of single-crystal diffraction patterns, locally forming partial powder-diffraction (Debye) rings. The position and contrast of most of the visible diffraction spots and rings can be interpreted as Fe3C reflections (e.g. 121, 031 the strongest, 230, etc.).

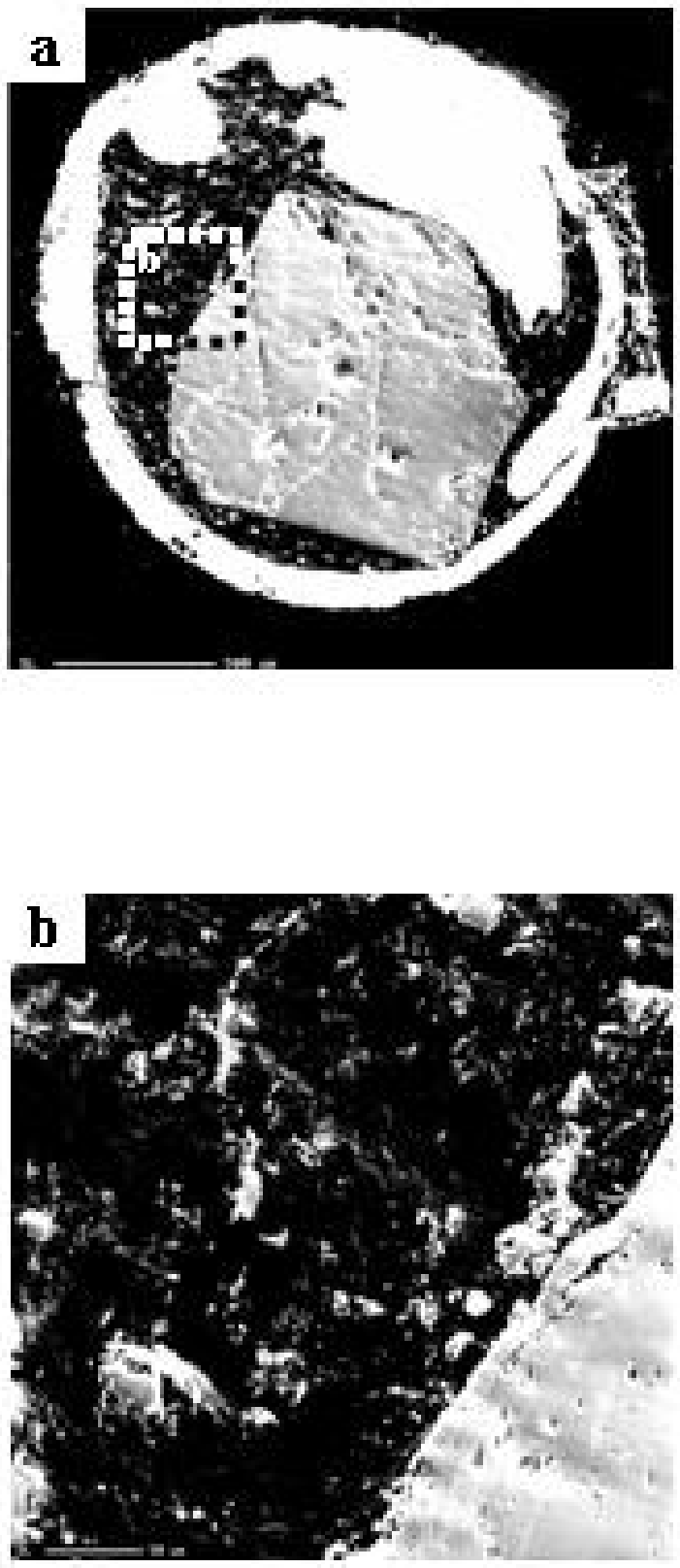

**Fig. 4** SEM image of the experimental product from a 12 hours long experiment with graphite, obtained at 6 GPa and 1400.

**a.)** The recrystallized San Carlos olivine is in the middle of the rhenium capsule and is surrounded by graphite.
**b.)** An enlarged part of the image. The material transported into the graphite was identified as garnet, pyroxene, spinel, and possibly quenched melt.

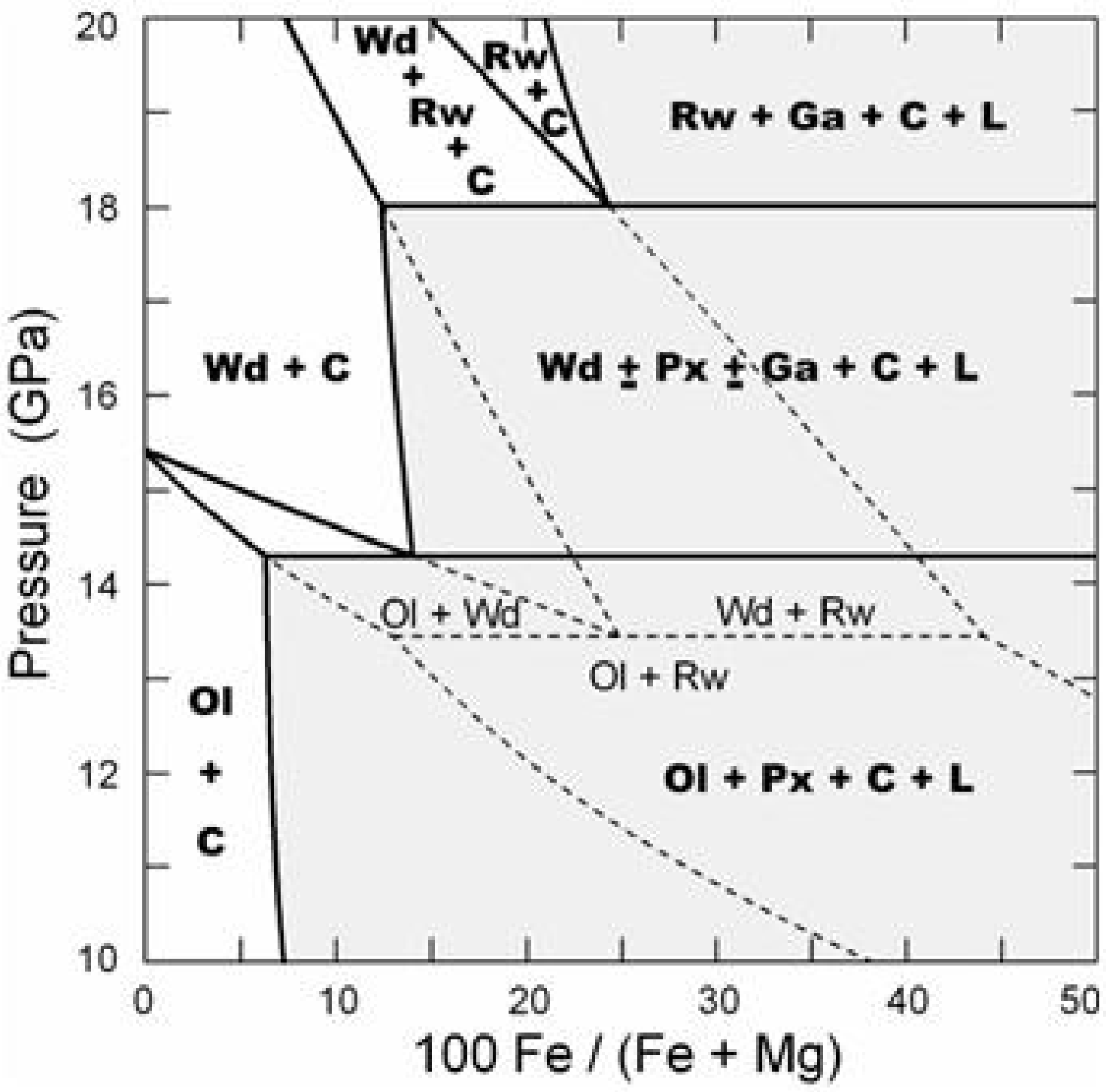

**Fig. 5** Pressure-composition phase diagram for the system, showing the subsolidus phase relations (dashed) calculated along the geotherm in Fig. 6 using the experimental data of Katsura and Ito (1989), and modified to include the effects of carbonate melting (solid). Only the shaded areas remain above the solidus
Symbols: C=diamond; Ga = garnet; L = liquid/melt; Ol = olivine;
perovskite; Px = pyroxene; Rw = ringwoodite; Wd = wadsleyite

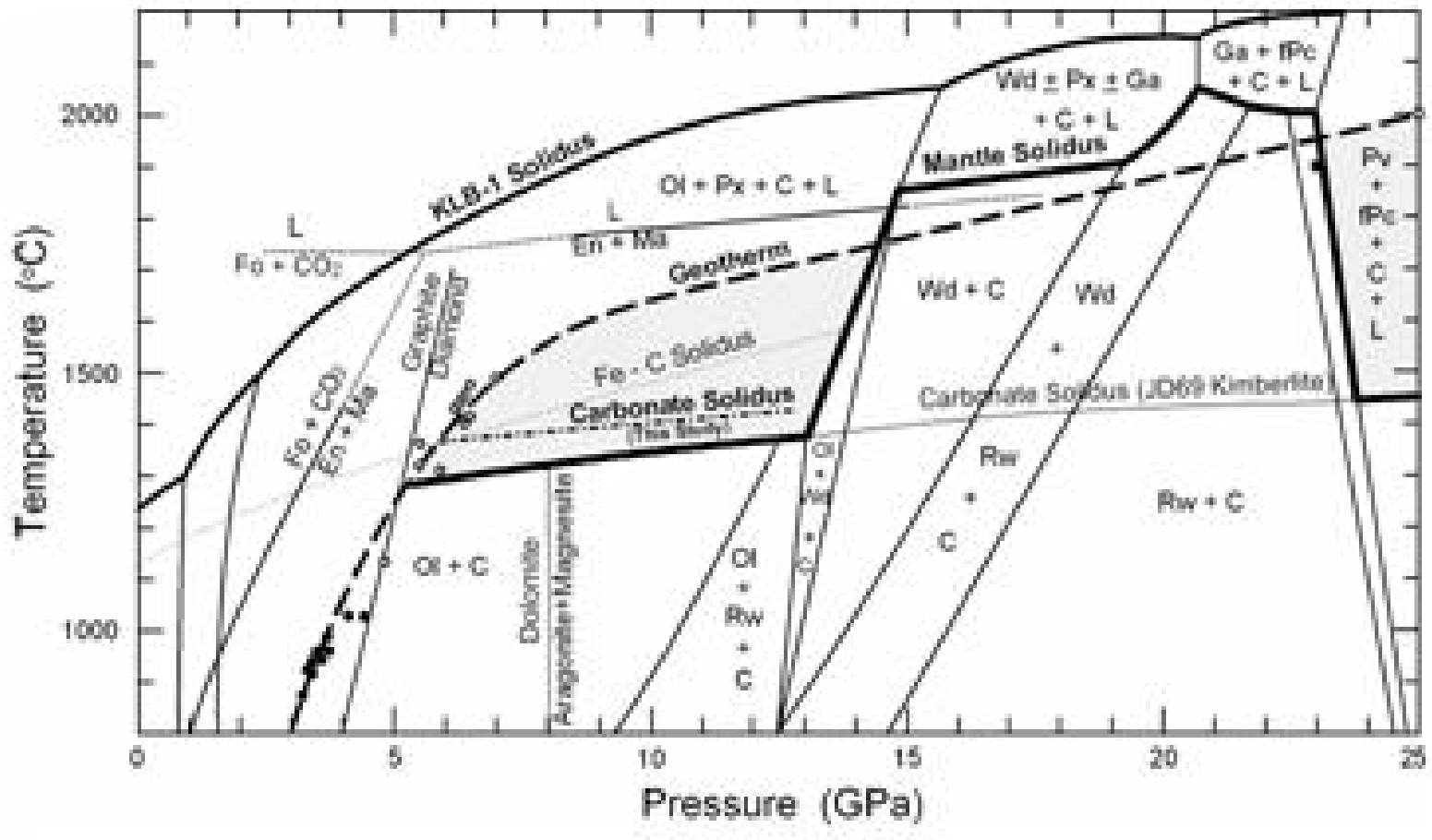

**Fig. 6** Temperature-pressure phase diagram for the composition . The geotherm was constrained at low pressures by the thermobarometry (Gasparik 2000) applied to mantle xenoliths from Northern Lesotho(Nixon and Boyd 1973), and at the lower-mantle pressures by the thermobarometry (Gasparik and Hutchison 2000) applied to the NaPx-En (Wang and Sueno 1996) and the Type III (Hutchison 2000) inclusions in diamonds. Symbols are the same as in Fig. 5; fPc = ferropericlase, (Mg,Fe)O